# Graphene allotropes: stability, structural and electronic properties from DF-TB calculations

**A.N. Enyashin, A.L. Ivanovskii**[*]

*Institute of Solid State Chemistry, Ural Branch of the Russian Academy of Sciences, 620990 Ekaterinburg, Russia*

A B S T R A C T

Using the density-functional-based tight-binding method we performed a systematic comparative study of stability, structural and electronic properties for 12 various types of graphene allotropes, which are likely candidates for engineering of novel graphene-like materials.

[*] Corresponding author. Fax: +7(343)3744495
*E-mail address:* ivanovskii@ihim.uran.ru (A.L. Ivanovskii).

## 1. Introduction

Graphene, a single-atom-thick flat two-dimensional (2D) carbon material [1] consisting of a monolayer of carbon atoms in a honeycomb array exhibits unusual electronic [2] and magnetic [3] properties, such as high carrier mobility and ambipolar effect, represents a unique metallic membrane, and therefore is considered as a revolutionary material for future generation of high-speed electronic, radio frequency logic devices, thermally and electrically conductive reinforced composites, sensors, transparent electrodes *etc* [2-6].

The above reasons and future application prospects of graphene-like materials have triggered much interest and stimulated numerous studies of further ways of controllable manipulation of the atomic and electronic properties of graphene. One of such approaches is introduction of atomic-scale defects such as various topological defects, atomic vacancies or dislocations and grain boundaries into the graphene network, which may lead to local or non-local disorder in the perfect graphene lattice and to loss of symmetry. Another approach employs structural deformation of grapheme by various strains or convolutions. A very promising approach of nano-engineering is the so-called chemical functionalization, where modification of atomic and electronic properties of graphene can be achieved by partial replacement of carbon atoms in 2D network by various dopants; by decoration of various defects in graphene with dopants; by covalent or non-covalent interactions of atoms or molecules with graphene network and so on [7,8]. New electronic properties (as compared with free standing graphene networks) arise as a result of formation of multilayer graphene materials or interaction of graphene with various substrates.

Let us note that in all of the above approaches the basic material is pristine graphene formed by carbon $sp^2$ atoms arranged in a honeycomb structure of hexagons $C_6$.

An alternative way of creation of graphene-like materials is the search for single-atom-thick periodic carbon networks constructed from non-$C_6$ carbon polygons, which can include also other groups of carbon atoms, for example, acetylenic linkages. Some models of such non-$C_6$ flat carbon networks, as well as possible synthetic pathways are discussed earlier, reviews [9,10].

In this work, by means of the density-functional-based tight-binding (DF-TB) method, we have performed a systematic theoretical study of stability, structural and electronic properties for 12 various types of two-dimensional planar carbon networks, which may be considered as graphene allotropes, and which are likely candidates for engineering of novel graphene-like materials.

## 2. Models and computational details

The atomic models of the examined carbon networks (*2-13´*) together with graphene (*1*) are depicted in Fig. 1. We began our simulations of possible graphene allotropes with the so-called *pentaheptites* [11,12] - flat networks *2* and *3*, where hexagons $C_6$ of graphene network (*1*) are replaced by pentagons $C_5$ or heptagons $C_7$. This construction implies partial or complete introduction of Stone-Wales (SW) defects in graphene, which involves an in-plane 90° rotation of two carbon atoms with respect to the midpoint of the bond. As a result, four hexagons will be converted into two heptagons and two pentagons [11]. In other words, *pentaheptites* arise by simultaneous application of bond rotation to four-hexagon pyrene units over the whole graphene plane [12]. The examined structures *2* and *3* differ by periodicity of distributions of $C_5$ and $C_7$ units.

The planar network *4* represents an "intermediate" case between graphene and *pentaheptites*, including hexagons $C_6$, pentagons $C_5$, and octagons $C_8$. Similar planar structures are known as *haeckelites* [13].

The next group of carbon flat networks (*5* and *6*) includes the so-called *graphynes* [14,15] consisting of hexagons $C_6$ interconnected by linear carbon chains. These networks can be constructed by replacing one-third of carbon-carbon bonds in graphene by acetylenic linkages (-C≡C-, structure *5*). By increasing the length of linear carbine-like atomic chains between carbon hexagons it is possible to design other related networks (termed also as *graphdiynes* [16], structure *6*). Replacement of ⅔ of carbon-carbon bonds by carbine-like chains leads to the structure (*9*) formed by distorted hexagons, whereas complete replacement of carbon-carbon bonds by carbine-like chains in graphene gives rise to the related graphene-like structure (*10*), which can be named as *supergraphene*.

Network *7* is composed of equal numbers of squares and octagons [17], whereas a more complicated network *8* is organized by squares $C_4$ and decagons $C_{10}$. The hypothetical polycyclic network *11* is composed by $C_3$ and $C_{12}$ cycles.

Finally, networks *12* and *13* (so-called *squarographenes* [18]) are formed by distorted hexagons and regular squares (*12*) or by undistorted hexagons and rhombuses (*13*).

All the considered allotropes can be divided into three groups using Heimann's et al. [19] classification based on valence orbital hybridization (n). In this way, the majority of the examined allotropes (*2, 3, 4, 7, 8,* and *11*), like graphene, belong to the systems with "pure" $sp^2$ hybridization type (n = 2). On the other hand, networks *5, 6, 9*, and *10*, which include fragments of $sp^1$ carbine-like chains, adopt a "mixed" ($sp^2 + sp^1$) hybridization type (1 < n < 2), whereas network *12* and the starting network *13*, see also below, comprise three-fold ($sp^2$) and four-fold ($sp^3$) coordinated carbon atoms and belong to another "mixed" ($sp^2 + sp^3$) hybridization type (2 < n < 3).

All the calculations were performed within the density-functional-based tight-binding method (DF-TB) [20-22], which was earlier used for simulations of various carbon structures and was found to describe these systems in reasonable agreement with experimental data and high-level theoretical methods, see [23-28].

### 3. Results and discussion

As the first step, full geometry optimization was carried out for all the above 2D carbon networks. The self-consistent calculations were considered to be converged when the difference in the total energy did not exceed $10^{-4}$ eV as calculated at consecutive steps, and the forces between atoms were close to zero. We found that almost all of the considered systems (*1-12*) keep their initial flat structures after relaxation - except network *13*, which transforms into a new much more stable flat network formed by infinite strips of hexagons interconnected by carbine-like atomic chains (*13´*).

For ideal graphene, the DF-TB calculations of the structural and electronic parameters demonstrate that the main results, namely the obtained C-C distances (2.47 Å) and the semimetal-like band structure type (where the $\pi - \pi^*$ band crossing at $E_F$ is strictly gapless, Fig. 2), agree well with the data of more sophisticated *ab-initio* calculations, reviews [2,8]. The optimized structural parameters for the other systems *2-13´* are presented in Table 1.

Let us now focus on comparative stability of the examined systems and the ideal graphene network. For this purpose, the differences in the total energies ($\delta E$, per carbon atom) between the examined allotrope $E_{tot}^{allotr.}$ and pristine graphene $E_{tot}^{graph.}$: $\delta E = (E_{tot}^{allotr.} - E_{tot}^{graph.})$ were calculated. The results (Table 2) show that $\delta E$ for all the considered allotropes are positive; this means that they are meta-stable in comparison with graphene.

The parameters $\delta E$ presented in Table 2 allow us to make the following conclusions. First, the formation of flat networks of ($sp^2 + sp^3$) hybridization type (structures *12* and *13*) and of allotropes *9, 10,* and *11*, which are constructed from fragments of $sp^1$ carbine-like chains (*9* and *10*) or *11*, composed by $C_3$ and $C_{12}$ cycles, will be most unfavorable. Second, from Table 2 it is seen that in comparison with the above networks, the combination of $sp^1$ carbine-like chains with hexagons $C_6$ leads to stabilization of the corresponding structures *5*, *6,* and *13´*; however, as the $sp^1$ chains grow in length, the stability of the network decreases (*6 versus 5*). Finally, the most stable structures (which adopt the minimal $\delta E$ values) are allotropes *2-4*, *i.e. pentaheptite-* and *haeckelite*-like flat networks of "pure" $sp^2$ hybridization type formed by periodically distributed $C_n$ cycles, where n = 5, 6, 7 and 8. Hence, it is possible to speculate that synthesis of these flat 2D carbon structures is most probable, whereas inclusion other carbon cycles or $sp^1$ chains in the corresponding networks should be considered as a negative factor. The preconditions for synthesis of some of the considered allotropes are discussed in Refs. [9-18].

Now we shall discuss the electronic band structure of the examined allotropes focusing on their conducting properties. As can be seen from Figs. 2, 3 and Table 1, all the examined flat networks can be divided into three groups:

(i) gapless semi-metals (graphene *1* and *supergraphene 10*); (ii) metallic-like systems (*2, 3, 7, 9, 11, 12,* and *13*) and (iii) semiconductors (*4, 5, 6, 8,* and *13´*).

Thus, the semiconducting behavior was found for *haeckelite*-like network (*4*), for all the considered types of *graphyne*-like networks (*5* and *6*), as well as for polycyclic networks *8* and *13´*. Therefore our preliminary conclusion is that for the opening of the band gap, flat carbon networks should contain at least the combinations of $C_5 + C_6 + C_8$ units (*4*), or $C_4 + C_5$ units (*8*), or fragments of carbine-like chains + hexagons $C_6$ (*5, 6* and *13´*). All other types of the examined structures are metallic-like or gapless semi-metals.

Besides, let us note that among the semiconducting allotropes there are only systems of "pure" $sp^2$ - (*4* and *8*) or "mixed" ($sp^2 + sp^1$) hybridization types (*5, 6,* and *13´*), whereas all the considered ($sp^2 + sp^3$)-like allotropes are metallic-like. The density of states near the Fermi level for these metallic-like structures adopt maximal values for allotropes *7, 11,* and *12*, see Fig. 3.

In addition, the peculiarities of distributions of the electronic states near the Fermi level can be examined using Γ-point isosurfaces of squared orbitals $\psi^2$ for HOMO and LUMO, see Fig. 4. It is visible that depending on the spectrum around zero energy all the examined structures can be divided into two main groups, namely symmetric and asymmetric. So, among semiconducting allotropes (*4, 5, 6, 8,* and *13´*) the top of valence π bands (Γ-point isosurfaces of $\psi^2$ for HOMO) and the bottom of conduction π* bands (Γ-point isosurfaces of $\psi^2$ for LUMO) arise mainly from comparable contributions of states of the same carbon atoms (for *5, 6,* and *8*) or the same separate atomic groups (for *4*), and these systems are of symmetric type. On the contrary, network *13´* belongs to the asymmetric type: the top of the valence bands is formed by contributions of the $sp^1$ carbon atoms from carbine-like chains, whereas the bottom of the conduction bands arises mainly from the contributions of $sp^2$ carbon atoms forming hexagons $C_6$. A similar picture is demonstrated by metallic-like allotropes (*2, 3, 7, 9, 11, 12,* and *13*), which can be divided also into symmetric (*7* and *9*) and asymmetric (*2, 3, 11, 12,* and *13*) systems.

## 4. Conclusions

The main purpose of this work is to examine the structural, electronic properties and comparative stability for the family of 12 alternative two-dimensional flat carbon networks, which are likely candidates for engineering of novel graphene-like materials.

As a result, their stability hierarchy was considered. We found that the most stable structures are flat networks of "pure" $sp^2$ hybridization type formed by periodically distributed $C_n$ cycles, where n = 5, 6, 7 and 8, whereas inclusion of other carbon cycles or $sp^1$ chains into the networks should be considered as a negative factor. The examined allotropes demonstrate a rich set of electronic spectra of metallic, gapless semi-metallic and semiconducting types. Moreover, their band structure parameters (band gaps or near-Fermi density of states), as well as the composition of the near-Fermi states (HOMO-LUMO) vary in a wide range. The latter may be symmetric or asymmetric, *i.e.* formed by comparable contributions from states of the same carbon atoms or atomic groups, or by contributions of non-equivalent carbon atoms, opening thereby a new road towards search of likely candidates for engineering of novel graphene-like materials.

One of the interesting results of our study concerns the data for *supergraphene*, which may be considered as the second example (along with graphene) of gapless semi-metals, where (similarly to graphene) the electronic spectrum close to the Fermi energy has a conical form mimicking the dispersion of a relativistic massless Dirac particle. However, unlike graphene, the corresponding states are formed by two types of non-equivalent carbon atoms of $sp^2$ and $sp^1$ hybridization types.

Finally, here we focused entirely on the electronic properties and comparative stability of the "pure" flat carbon nerworks as possible graphene allotropes. In turn, it is possible to assume that the examined allotropes may be used as initial systems for further modification in the same manner as the above mentioned ''conventional'' graphene, i.e. by various topological defects, chemical functionalization etc. This would open up interesting prospects for the design of novel members of the family of graphene-like materials.

**References**

[1] K.S. Novoselov, A.K. Geim, S.V. Morozov, D. Jiang, Y.Zhang, S.V. Dubonos, I.V. Grigorieva, A.A. Firsov, Science 306 (2004) 666.
[2] A.H. Castro Neto, F. Guinea, N.M.R. Peres, K.S. Novoselov, A.K. Geim, Rev. Mod. Phys. 81 (2009) 109.
[3] E. Kan, Z.Y. Li, J.L.Yang, NANO, 3 (2008) 433.
[4] P. Avouris, Z.H. Chen, V. Perebeinos, Nature Nanotech. 2 (2007) 605.
[5] M. Dragoman, D. Dragoman, Prog. Quantum Electronics 33 (2009) 165.
[6] A.H. Castro Neto, Mater. Today 13 (2010) 1.
[7] M.J. Allen, V.C. Tung, R.B. Kaner, Chem. Rev. 110 (2010) 132.
[8] D.W. Boukhvalov, M.I. Katsnelson, J. Phys.: Condens. Matter 21 (2009) 344205.
[9] A.L. Ivanovskii, Russ. J. Inorg. Chem. 50 (2005) 1408.
[10] V.V. Pokropivny, A.L. Ivanovskii, Uspekhi Khimii 77 (2008) 899.
[11] V.H. Crespi, L.X. Benedict, M.L. Cohen, S.G. Louie, Phys. ReV. B 53 (1996) R13303.
[12] M. Deza, P.W. Fowler, M. Shtogrin, K. Vietze, J. Chem. Inf. Comput. Sci. 40 (2000) 1325.
[13] H. Terrones, M. Terrones, E. Hernández, N. Grobert, J-C. Charlier, P. M. Ajayan, Phys. Rev. Lett. 84 (2000) 1716.
[14] R.H. Baughman, H. Eckhardt, M. Kertesz, J. Chem. Phys. 87 (1987) 6687.
[15] N. Narita, S. Nagai, S. Suzuki, K. Nakao, Phys. Rev. B 58 (1998) 11009.
[16] M.M. Haley, S.C. Brand, J.J. Pak, Angew. Chem. Int. Ed. 36 (1997) 836.
[17] H. Zhu, A.T. Balaban, D.J. Klein, T.P. Zivkovic, J. Chem. Phys. 101 (1994) 5281.
[18] M.J. Bucknum, E.A. Castro, Solid State Sci. 10 (2008) 1245.
[19] R.B. Heimann, S.E. Evsyukov, Y. Koga, Carbon 53, (1997) 1654.
[20] D. Porezag, T. Frauenheim, T. Köhler, G, Seifert, R. Kaschner, Phys. Rev. B, 51, 12947 (1995).

[21] T. Frauenheim, G. Seifert, M. Elstner et al., J. Phys.: Condens. Matter, 14, 3015 (2002).
[22] G. Seifert, J. Phys. Chem. A 111 (2007) 5609.
[23] P.W. Fowler, T. Heine, K.M. Rogers, J.P.B. Sandall, G. Seifert, F. Zerbetto, Chem. Phys. Lett., 300, 369 (1999).
[23] H. Hermann, K. Zagorodniy, A. Touzik, M. Taut, G. Seifert, Microel. Engineering, 82, 387 (2005).
[24] A. Kuc, G. Seifert, Phys. Rev. B 74 (2006) 214104.
[25] A.N. Enyashin, A.L. Ivanovskii, JETP Lett. 86 (2007) 537.
[26] A.N. Enyashin, A.L. Ivanovskii, Phys. Rev. B 77 (2008) 113402.
[27] A.N. Enyashin, A.L. Ivanovskii, Chem. Phys. Lett. 473 (2009) 108.
[28] A. Kuc, T. Heine, G. Seifert, Phys. Rev. B 74 (2010) 085430.

**Table 1.** The optimized average C-C distances ($d$, in Å) and band gaps (BG, in eV) for graphene (*1*) and the examined graphene allotropes (*2-13´*) according to DF-TB calculations.

| system * | $d(sp^1\text{-}sp^1)$** | $d(sp^1\text{-}sp^2)$ | $d(sp^2\text{-}sp^2)$ | $d(sp^2\text{-}sp^3)$ | BG |
|---|---|---|---|---|---|
| *1* | - | - | 1.427 | - | 0 |
| *2* | - | - | 1.419 | - | 0 |
| *3* | - | - | 1.411 | - | 0 |
| *4* | - | - | 1.395 | - | 0.76 |
| *5* | 1.220 | 1.432 | 1.419 | - | 1.32 |
| *6* | 1.304 | 1.428 | 1.423 | - | 1.47 |
| *7* | - | - | 1.429 | - | 0 |
| *8* | - | - | 1.449 | - | 1.14 |
| *9* | 1.237 | 1.413 | 1.422 | - | 0 |
| *10* | 1.248 | 1.402 | - | - | 0 |
| *11* | - | - | 1.403 | - | 0 |
| *12* | - | - | 1.458 | 1.530 | 0 |
| *13* | - | - | 1.315 | 1.580 | 0 |
| *13´* | 1.236 | 1.426 | 1.446 | - | 0.70 |

\* see Fig. 1.
\*\* for carbon atoms of various hybridization types, see the text.

**Table 2**. The calculated lattice constants ($a$ and $b$, in Å) and relative energies ($\delta E$, eV/atom) for graphene (**1**) and the examined graphene allotropes (**2-13´**) according to DF-TB calculations.

| system * | lattice type | Z ** | lattice constants | δE *** | n**** |
|---|---|---|---|---|---|
| **1** | hexagonal | 2 | $a = 2.47$ | 0 | $sp^2$ |
| **2** | rectangular | 16 | $a = 4.87; b = 8.84$ | 0.308 | $sp^2$ |
| **3** | rectangular | 16 | $a = 5.70; b = 7.56$ | 0.323 | $sp^2$ |
| **4** | square-like | 28 | $a = 8.71$ | 0.397 | $sp^2$ |
| **5** | hexagonal | 36 | $a = 11.99$ | 0.695 | $sp^2 + sp^1$ |
| **6** | hexagonal | 54 | $a = 16.52$ | 0.867 | $sp^2 + sp^1$ |
| **7** | square-like | 4 | $a = 3.47$ | 0.707 | $sp^2$ |
| **8** | hexagonal | 12 | $a = 6.83$ | 0.828 | $sp^2$ |
| **9** | rectangular | 12 | $a = 7.03; b = 6.91$ | 0.928 | $sp^2 + sp^1$ |
| **10** | hexagonal | 8 | $a = 7.02$ | 1.035 | $sp^2 + sp^1$ |
| **11** | hexagonal | 6 | $a = 5.24$ | 1.386 | $sp^2$ |
| **12** | square-like | 5 | $a = 3.62$ | 2.077 | $sp^2 + sp^3$ |
| **13** | rectangular | 3 | $a = 2.72; b = 2.92$ | 2.914 | $sp^2 + sp^3$ |
| **13´** | rectangular | 6 | $a = 2.51; b = 6.96$ | 0.537 | $sp^2 + sp^1$ |

\* see Fig. 1.
\*\* numbers of C atoms in cell;
\*\*\* relative to $E_{tot}$ of graphene
\*\*\*\* hybridization types, see the text.

**FIGURES**

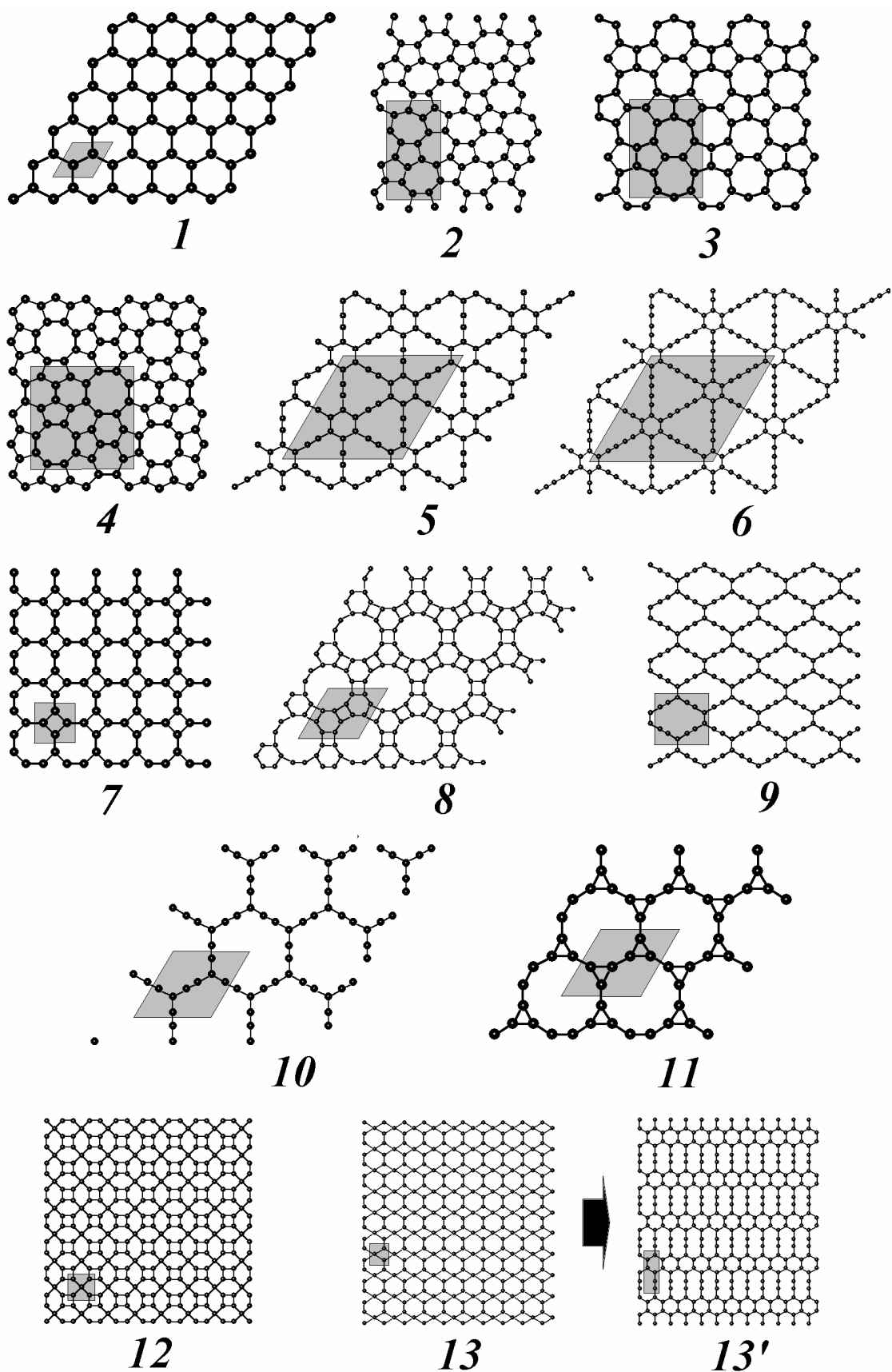

**Fig. 1.** Optimized atomic structures for the examined graphene allotropes (*13'*- see the text). Unit cells are painted.

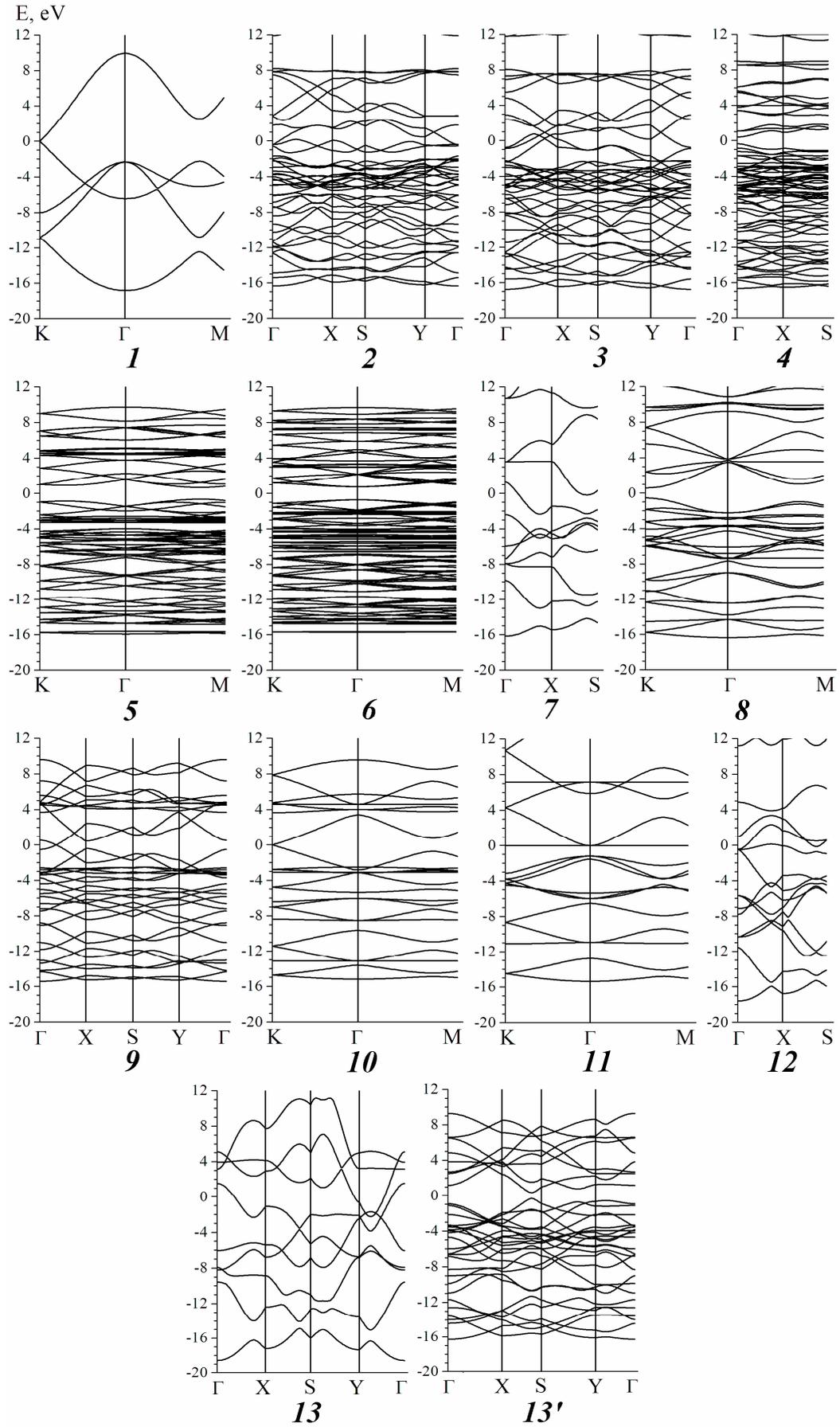

**Fig. 2.** Electronic band structures for the examined graphene allotropes (*1-13′*- see Fig. 1) according to DF-TB calculations. Fermi level is set to 0 eV.

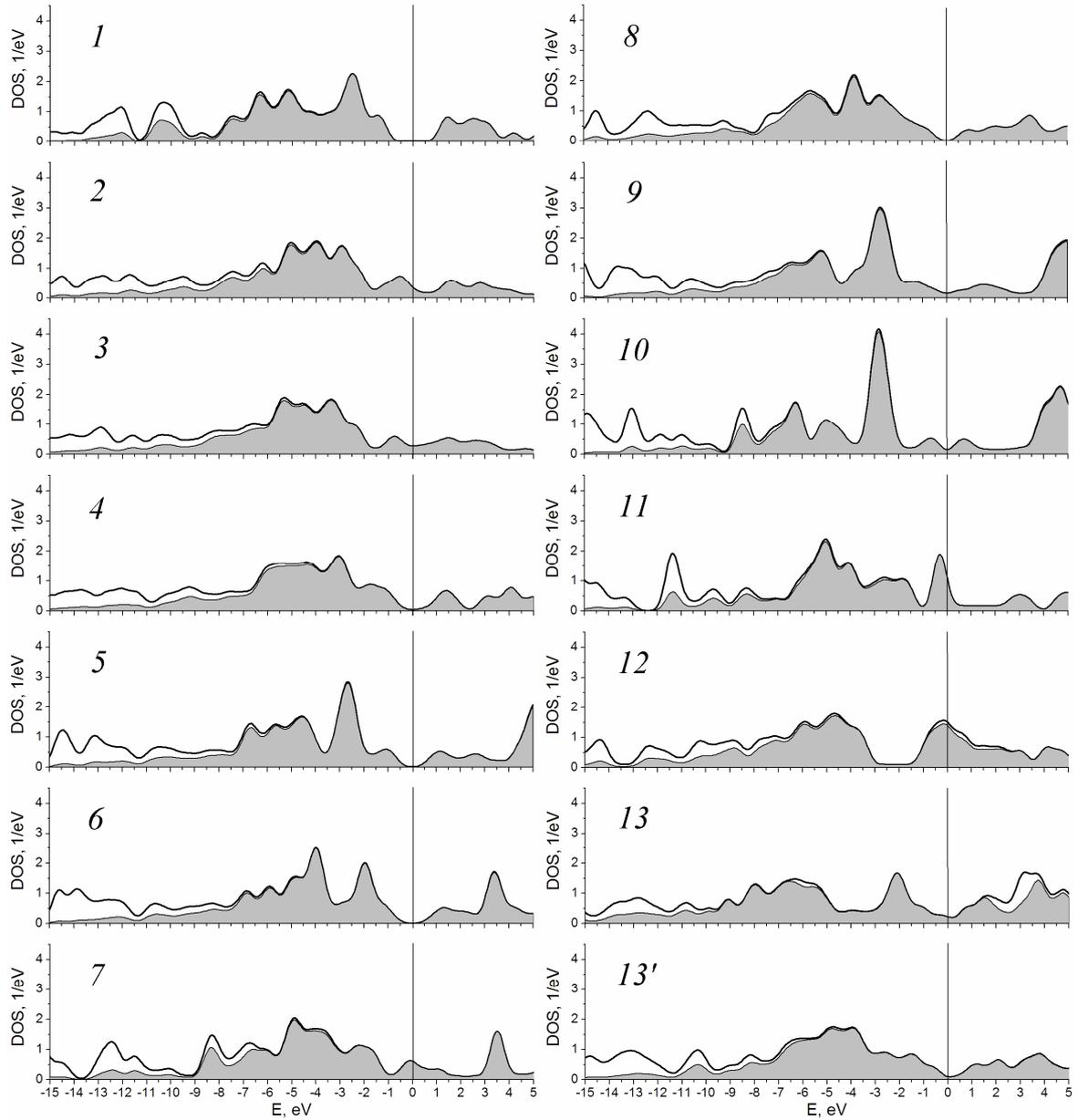

**Fig. 3.** Densities of states for the examined graphene allotropes (*1-13´* - see Fig. 1) according to DF-TB calculations. Fermi level is set to 0 eV. C 2*p* - states are painted in gray.

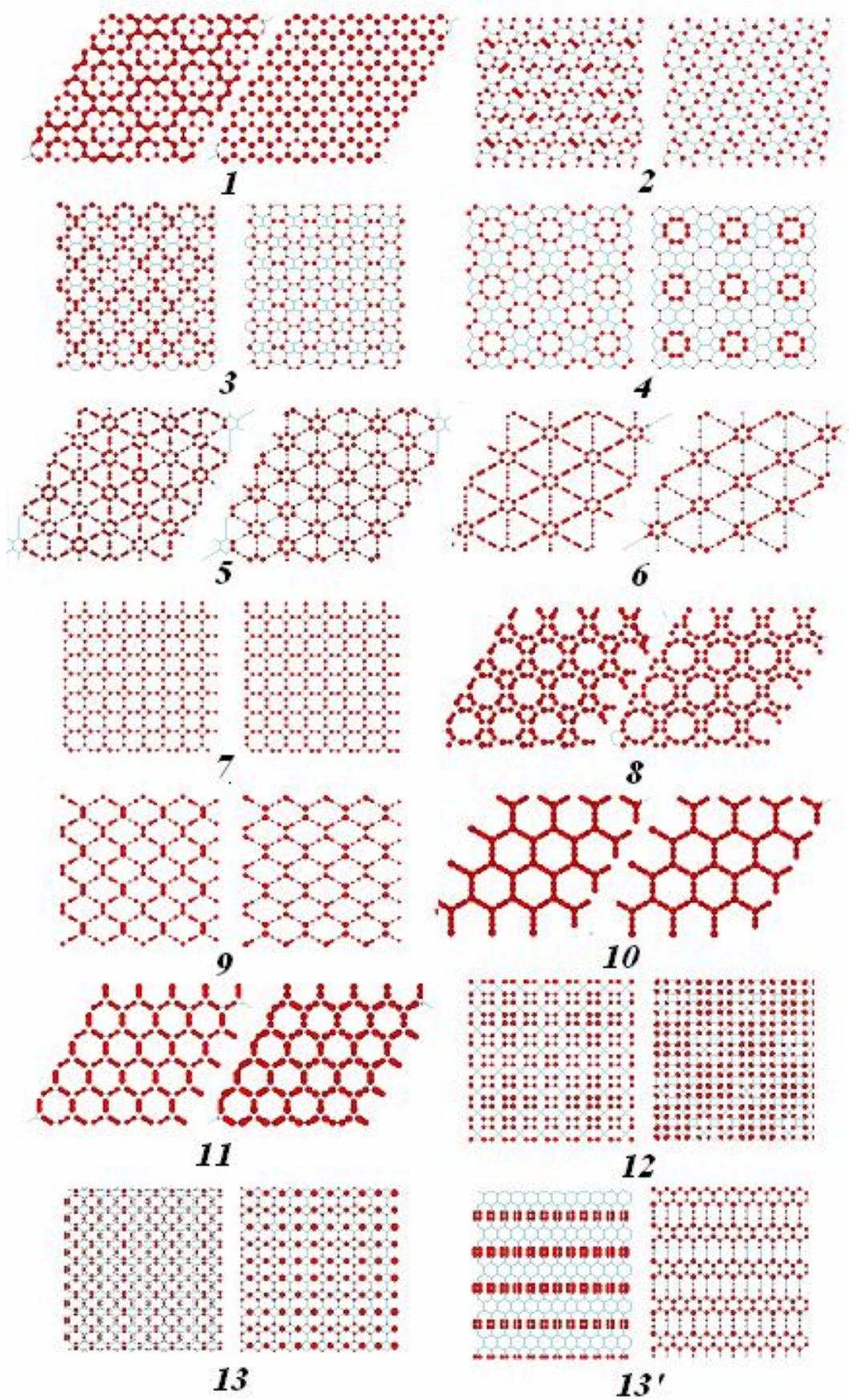

**Fig. 4.** (*Color online*) Γ-point isosurfaces of squared orbitals $\psi^2$ for HOMO (*left*) and LUMO (*right*) of the examined graphene allotropes (*1-13´* - see Fig. 1).